\documentclass[aps,pra,amsmath,amssymb,reprint,superscriptaddress,tightenlines]{revtex4-1} 

\usepackage{color}

\usepackage{afterpage}
\usepackage{graphicx}
\usepackage{amsmath}

\usepackage[pagebackref=false]{hyperref}

\begin{document}

\title{High-Fidelity Spin Measurement on the Nitrogen--Vacancy Center}

\author{Michael Hanks} \email[]{hanks@nii.ac.jp}
\affiliation{Department of Informatics, School of Multidisciplinary Sciences, Sokendai (The Graduate University for Advanced Studies), 2-1-2 Hitotsubashi, Chiyoda-ku, Tokyo 101-8430 Japan}
\affiliation{National Institute of Informatics, 2-1-2 Hitotsubashi, Chiyoda-ku, Tokyo 101-8430, Japan}
\author{Michael Trupke}
\affiliation{Quantum Optics, Nanophysics and Information Group, Faculty of Physics, University of Vienna, Boltzmanngasse 5, Vienna A-1090, Austria}
\affiliation{Vienna Center for Quantum Science and Technology, Atominstitut, TU Wien, Stadionallee 2, 1020 Vienna, Austria}
\author{J{\"o}rg Schmiedmayer}
\affiliation{Vienna Center for Quantum Science and Technology, Atominstitut, TU Wien, Stadionallee 2, 1020 Vienna, Austria}
\author{William J. Munro}
\affiliation{NTT Basic Research Laboratories, NTT Corporation, 3-1 Morinosato-Wakamiya, Atsugi, Kanagawa 243-0198, Japan}
\affiliation{NTT Research Center for Theoretical Quantum Physics, NTT Corporation, 3-1 Morinosato-Wakamiya, Atsugi, Kanagawa 243-0198, Japan}
\affiliation{National Institute of Informatics, 2-1-2 Hitotsubashi, Chiyoda-ku, Tokyo 101-8430, Japan}
\author{Kae Nemoto}
\affiliation{National Institute of Informatics, 2-1-2 Hitotsubashi, Chiyoda-ku, Tokyo 101-8430, Japan}
\affiliation{Department of Informatics, School of Multidisciplinary Sciences, Sokendai (The Graduate University for Advanced Studies), 2-1-2 Hitotsubashi, Chiyoda-ku, Tokyo 101-8430 Japan}

\date{\today}

\begin{abstract}
{Nitrogen-vacancy (NV) centers in diamond are versatile candidates for many quantum information processing tasks, ranging from quantum imaging and sensing through to quantum communication and fault-tolerant quantum computers. Critical to almost every potential application is an efficient mechanism for the high fidelity readout of the state of the electronic and nuclear spins. Typically such readout has been achieved through an optically resonant fluorescence measurement, but the presence of decay through a meta-stable state will limit its efficiency to the order of 99\%. While this is good enough for many applications, it is insufficient for large scale quantum networks and fault-tolerant computational tasks. Here we explore an alternative approach based on dipole induced transparency (state-dependent reflection) in an NV center cavity QED system, using the most recent knowledge of the NV center's parameters to determine its feasibility, including the decay channels through the meta-stable subspace and photon ionization. We find that single-shot measurements above fault-tolerant thresholds should be available in the strong coupling regime for a wide range of cavity--center cooperativities, using a majority voting approach utilizing single photon detection. Furthermore, extremely high fidelity measurements are possible using weak optical pulses.}
\end{abstract}

\maketitle

\section{Introduction} \label{sec:introduction}

The twentieth century saw the discovery of quantum mechanics, a fundamental branch of physics concerning systems such as atoms and molecules that can exist in a `quantum superposition' of different states \cite{Schrodinger26,vanderWaerden68}. This has had a profound impact on our understanding of our natural world. The recent loophole-free Bell inequality tests \cite{Giustina15,Shalm15,Hensen15,Hensen16} have shown nonlocal correlations that cannot be described solely by classical physics. Quantum principles such as superposition and entanglement \cite{Einstein35,Horodecki09} have no classical counterparts and are now known to allow a twenty first century technological paradigm shift. Quantum technologies promise unparalleled performance in computation \cite{Deutsch92,Simon94,Shor94,Grover96}, the simulation of physical systems \cite{Feynman82,Cirac12,Georgescu14}, secure communication \cite{Wiesner83,Bennett84} and metrology \cite{Giovannetti04,Giovannetti06,Paris09,Giovannetti11,DemkowiczDobrzanski12,Toth14,Taylor14}. Many physical systems, including ion-traps \cite{Cirac95,Wineland98,Cirac00,Kielpinski02,Leibfried03,Porras04,Haffner05,Blatt08, Haffner08,Kim10,Blatt12}, superconducting circuits \cite{Blais04,Wallraff04,You05,Hofheinz08,You11,Houck12,Devoret13,Xiang13,Barends14,Kelly15}, quantum dots \cite{Loss98,Burkard99,Imamoglu99,Benson00,Michler00,Petta05,Hennessy07,Hanson07,Kairdolf13}, linear optics \cite{Knill01,Kok07,OBrien07}, donor spins in silicon \cite{Kane98,Vrijen00,Morton08,Pla12,Tyryshkin12,Steger12,Pla13,Zwanenburg13} and nitrogen--vacancy centers among other defects \cite{Davies92,Zaitsev00,Jelezko06,Wrachtrup06,Greentree08,Weber10,Aharonovich11,Koehl11,Aharonovich11,Dzurak11,Gordon13}, have been proposed to realize such technologies. Key experiments have been performed demonstrating basic required operations \cite{Brouri00,Gaebel06,Togan10,Robledo11a,Maurer12,Zhou14,Rong15}, and small-scale applications have also been demonstrated \cite{Kurtsiefer00,Dutt07,Babinec10}.

As we move to larger-scale, more complex devices, whether for quantum communication, computation or simulation tasks, it will be essential to perform some form of error detection or correction. Quantum error correction demands that unitary operations and projective measurements achieve operational fidelities greater than a code-dependent \emph{accuracy threshold} (for instance the nearest-neighbor interactions based \emph{surface code} \cite{Dennis02} has a conservative threshold of $99.4$\% \cite{Stephens14}).  Resource scalability necessitates an order of magnitude greater than this, suggesting the practical threshold $99.9$\%, though these requirements may be lower for quantum communication schemes. It has been challenging to realize projective measurements with this accuracy in most physical systems.

\begin{figure*}[t]
\includegraphics[width=0.8\textwidth]{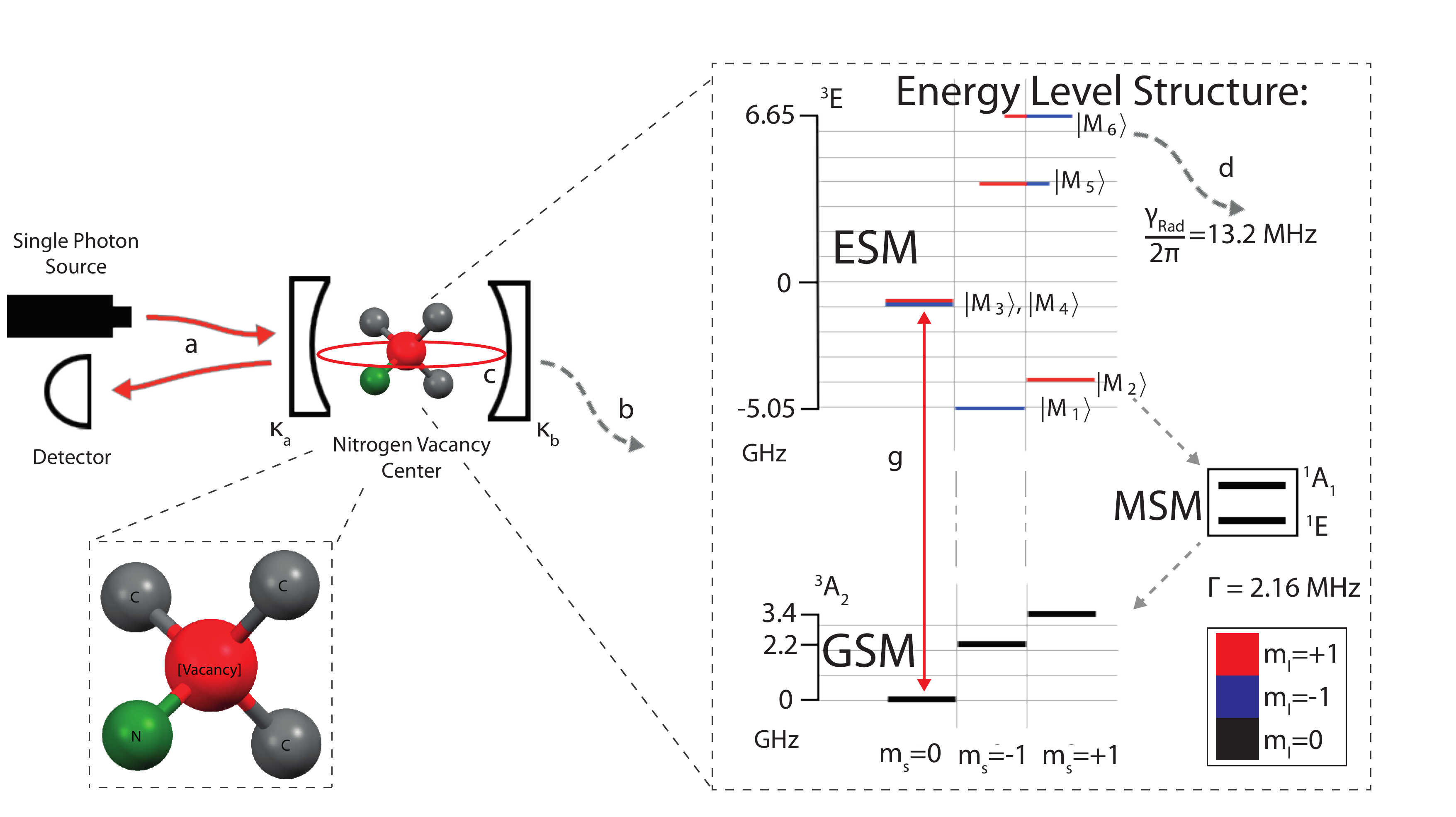}
\caption{{\bf (Left)} Schematic illustration for single-photon measurement of the NV$^-$ electron spin state based on dipole-induced transparency. The CQED system is driven by a single-photon source while detection in the reflected mode is used to infer information about the state of the NV$^-$ center. The NV center is composed of an adjacent nitrogen atom (green) and lattice vacancy (red) in a tetrahedral carbon lattice (gray). {\bf (Right)} The electronic energy level structure of the NV$^-$ center, at low temperature under a B$_{z}=20$ mT external magnetic field set along the NV axis. The energy levels are represented by bold horizontal lines and grouped into three subspaces: \textit{ground state manifold (GSM)}, \textit{excited state manifold (ESM)} and \textit{meta-stable state manifold (MSM)}. Allowed radiative transitions between the ground and optically excited states are horizontally segmented, with overlap indicating relative spin composition of the energy eigenstates. The optically excited states contain non-zero orbital angular momentum components $+1$ (red) and $-1$ (blue). Dashed arrows represent decay paths not resulting in reflection of a photon incident on the cavity (they represent transmission into the cavity mode, $c$, through the second cavity mirror (into $b$), spontaneous emission into the free field ($d$) and decay to a meta-stable state). The solid double-sided arrow represents the zero-phonon line transition in resonance with the optical cavity. }
\label{fig:full_structure}
\end{figure*}

The negatively charged nitrogen--vacancy center (NV$^-$) in diamond is certainly an interesting candidate system for quantum technologies, with potential applications ranging from quantum metrology \cite{Taylor08,Maze08,Balasubramanian08,Acosta09,Pham11,Dolde11,Maletinsky12,Barry16} to quantum communication \cite{Childress05,Childress06,Childress13}, as well as simulation and computation  \cite{Wrachtrup01,Nizovtsev05,Trajkov05,Ladd10,Weber10,Yang11,Childress13,Casanova16}. The NV$^-$ center contains electronic and nuclear spin components, with both optical and microwave transitions, and remains stable even at room temperature \cite{Doherty13}. At low temperature (4 - 8 K) the electronic spin coherence times are on the order of $1$--$100$ milliseconds \cite{Yang16,Balasubramanian09,McLellan16}. If the NV center is going to be used for tasks related to communication or computation, it needs to perform four core tasks:
\begin{itemize}
	\setlength{\itemsep}{1pt}
	\setlength{\parskip}{0pt}
	\setlength{\parsep}{0pt}
	\item Coherent manipulation of electronic and nuclear spins,
	\item Entanglement generation between remote electronic spins,
	\item Entanglement swapping between the electronic and nuclear spin states within an NV center, enabling the storage of remote entanglement in the nuclear spins,
	\item Measurement of the electronic and nuclear spin states.
\end{itemize}
Though local operations (the coherent manipulation) of single centers has been performed with fidelities exceeding $0.999952(6)$ \cite{Rong15}, sample fabrication and high fidelity measurements are technological challenges yet to be overcome \cite{Greentree06,Gupta16}. 

The typical measurement scheme for an NV$^-$ center is based on a cycling transition between the $m_s=0$ levels in the ground and excited state manifolds. The presence of decay channels from the optically excited manifold through a meta-stable state is expected to limit the efficiency of this scheme to order 99\%. The efficiency is therefore likely to be below the threshold required for larger-scale communication and computational tasks, though recent work utilizing repeated initialization of the electronic state between measurement trials \cite{Jiang09} may prove to circumvent this limitation in cases not related to the generation of entanglement. Cavity-enhanced transitions have been exploited in Purcell-enhanced fluorescence measurement schemes \cite{Wolters10,Faraon11,Faraon12} and in emission-based entanglement schemes \cite{Lim05,Barrett05,Jiang07}.

Recently an alternative approach based on dipole-induced transparency has been  proposed \cite{Duan04,Young09,Koshino12,Nemoto14,Sun16} wherein  the state of the NV center changes the resonance properties of an optical cavity. More specifically, if the electronic spin in its $\left|0\right>_{m_{s}}$ state, an incident photon is reflected from the cavity, while for the $\left|+1\right>_{m_{s}}$ electronic spin state the photon enters the cavity where it is scattered but not absorbed by the NV center. Detection of the reflected photon is thus a definite signature that the electronic spin was in the $\left|0\right>_{m_{s}}$ state, though the photon need not interact directly with the state of the NV center. In this work, we build upon the analysis of \cite{Nemoto14} and make use of recent improvements in the understanding of the NV's low-temperature optical characteristics \cite{Goldman15} to refine the quantification of the scheme's potential.

This paper is structured as follows: Section \ref{sec:the_nitrogen_vacancy_center} begins with an overview of the main properties of the nitrogen--vacancy center, followed in Section \ref{sec:photonic_readout_of_the_nitrogen_vacancy_center_state} by a description of the CQED system and the measurement scheme. Section \ref{sec:analysis} analyzes the measurement process in detail and estimates its performance, while Section \ref{sec:wcp} considers the effect of replacing the single photon source with weak coherent laser pulses. Finally, Section \ref{sec:discussion} summarizes the main conclusions of our analysis and briefly discusses some areas with the potential for improvement.

\section{The Nitrogen--Vacancy Center}  \label{sec:the_nitrogen_vacancy_center}

The negatively charged nitrogen--vacancy center in diamond consists of the nearest-neighbor pair of a subsitutional nitrogen atom with a lattice vacancy (green and red respectively in Figure \ref{fig:full_structure}).
Three dangling bonds from carbon atoms adjacent to the vacancy, two dangling bonds from the Nitrogen atom and an additional electron form the electronic structure of the center.
Isotopes of both nitrogen (14,15) and carbon (12,13) allow us to  tailor  the number and properties of the nuclear spins. $^{14}{\rm N}$ has a spin-1 nuclear spin while that of $^{15}{\rm N}$ is spin 1/2. Similarly, $^{12}{\rm C}$ has no nuclear spin while that of $^{13}{\rm C}$ is spin 1/2.  With isotopic engineering an NV center can be fabricated as one requires. For this article we consider an NV$^{-}$ center with the $^{15}$N isotope and no nearby $^{13}{\rm C}$ carbon atoms. This leads to the simplest NV center consisting of a spin-1/2 nucleus and an electronic structure (see Figure \ref{fig:full_structure}) is broadly classified by symmetry and/or multiplicity into three groups. These groups are the ground state manifold (GSM) $^{3}$A$_{2}$ (three non-degenerate states with a $180^{o}$-rotation symmetry about the principle axis), the optically excited state manifold (ESM) $^{3}$E (three doubly-degenerate states), and the meta-stable manifold (MSM) containing the singlets $^{1}$A$_{1}$, one non-degenerate state with both $180^{o}$-rotation and reflection symmetries, and $^{1}$E, one doubly-degenerate state.  The ground and optically excited states can be decomposed into two three-level subsystems. The ground states we refer to as the `electron spin' $S$, with quantum number $m_{s}$ (and the nuclear spin $I$, with quantum number $m_{n}$). This will distinguish it from optical excitation, which we refer to as a change in the `orbital angular momentum' $L$, with quantum number $m_{l}$.

\begin{table}[tbht]
\caption{\label{tbl:zeroFieldES} Relative energy levels E (and low temperature free lifetimes $\tau$ \cite{Goldman15,Collins83}) of the optically excited states, showing the polarization of dominant transitions at both zero and $20$ mT external fields.  Small mixing of spin states (below $1$\%) is not shown here. We also indicate the proportion of the spontaneous decay through a meta-stable (MS) state \footnote{Lifetimes and  decay percentages through the meta-stable subspace are estimated from the 0 mT values using the energy level mixing ratios. For the low-temperature excited-state Hamiltonian we are using parameters averaged between  those of \cite{Tamarat06,Batalov09}.}.}
\begin{ruledtabular}
\begin{tabular}{c | c | c | c | c | c | c}
		0 mT & E$_{2}$ & E$_{1}$ & E$_{x}$ & E$_{y}$ & A$_{1}$ & A$_{2}$ \\
		\hline
		E (GHz) & $-4.46$ & $-4.46$ & $-0.796$ & $-0.796$ & $3.98$ & $6.53$ \\
		$\tau$ (ns) & $7.5$ & $7.5$ & $12.1$ & $12.1$ & $5.1$ & $12.1$ \\
		MS decay & $38\%$ & $38\%$ & $0-1\%$ & $0-1\%$ & $54\%$ & $0-1\%$ \\		
          $\begin{matrix} m_{s}=-1   \\ m_{s}=0 \\ m_{s}=+1 \end{matrix}$
        & $\begin{matrix} \sigma_{-} \\ \: \\ \: \end{matrix}$
        & $\begin{matrix} \: \\ \: \\ \sigma_{+} \end{matrix}$
        & $\begin{matrix} \: \\ \sigma_{+} + \sigma_{-} \\ \: \end{matrix}$
        & $\begin{matrix} \: \\ \sigma_{+} - \sigma_{-} \\ \: \end{matrix}$
        & $\begin{matrix} \sigma_{+} \\ \: \\ \sigma_{-} \end{matrix}$
        & $\begin{matrix} \sigma_{+} \\ \: \\ -\sigma_{-} \end{matrix}$   \\
        \hline\hline
		20 mT & M$_{1}$ & M$_{2}$ & M$_{3}$ & M$_{4}$ & M$_{5}$ & M$_{6}$\\
		\hline
		E (GHz)  & $-5.05$ & $-3.87$ & $-0.82$ & $ -0.77$ & $3.87$ & $6.64$ \\
		$\tau$ (ns) & $7.5$ & $7.5$ & $12.1$ & $12.1$ & $5.2$ & $11.5$ \\
		MS decay & $38\%$ & $38\%$ & $0-1\%$ & $0-1\%$ & $52\%$ & $2-3\%$ \\		
		  $\begin{matrix} m_{s}=-1   \\ m_{s}=0 \\ m_{s}=+1 \end{matrix}$
		& $\begin{matrix} \sigma_{-} \\ \: \\ \: \end{matrix}$
		& $\begin{matrix} \: \\ \: \\ \sigma_{+} \end{matrix}$
		& $\begin{matrix} \: \\ \sigma_{-} \\ \: \end{matrix}$
		& $\begin{matrix} \: \\ \sigma_{+} \\ \: \end{matrix}$
		& $\begin{matrix} 0.83\sigma_{+} \\ \: \\ 0.56\sigma_{-} \end{matrix}$
		& $\begin{matrix}0.56 \sigma_{+} \\ \: \\ -0.83\sigma_{-} \end{matrix}$      
\end{tabular}
\end{ruledtabular}
\end{table}

The structure of this center can be described, under minor approximations \footnote{We ignore the axial hyperfine interaction in our Hamiltonian because it does not significantly change optical transition detunings and because its relative phase does not affect the fidelity of measurement in the spin basis. We have also removed cross-terms from the ground state Hamiltonian on the basis of their strength. These terms are the hyperfine spin-flip interaction, $A_{\perp} =3.65$ MHz \cite{Reichart05,Felton09} and the transverse strain field, $\delta_{x} \approx 1$ MHz \cite{Gruber97}. The energy gap, GHz, between the spin states they connect diminishes their effect. Moving to the dispersive-regime, the respective evolution terms can be approximated by $A^{2}_{\perp}t/(\text{GHz})$ and $\delta^{2}_{x}t/(\text{GHz})$. Measurement pulses are separated on timescales of the order $100$ ns; over the course of a single such pulse these quantities are expected to change state amplitudes to order $(10^{6})^{2}10^{-7}/10^{9}=10^{-4}$ and probabilities to order $10^{-8}$, which we can neglect}, by the Hamiltonian \cite{Doherty13}
\begin{align}
H_{NV} = H_{GSM}  \otimes \left|0\right>\left<0\right|_{m_{l}} + H_{ESM} 
\end{align}
where $H_{GSM}$ gives the structure of the ground state manifold and can be expressed as
\begin{eqnarray}
	H_{GSM} &=& \hbar D_{GSM} \left( S^{2}_{z} - \frac{2}{3} \right)+ \mu_{B} g^{||}_{GSM} S_{z} B_{z}\nonumber \\
	&&+ \mu_{N} g^{||}_{n} I^{(n)}_{z} B_{z}.
	\label{eqn:HamGSM}
\end{eqnarray}
Here $S$ and $I$ representing the usual electronic spin-1 and nuclear spin $\frac{1}{2}$ operators (with $S_z$ and $I_z$ being their respective z-components). The first term in $H_{GSM}$ represents an electronic spin zero field splitting of ${D_{GSM}}/{2\pi}=2.88$ GHz, while the second (third) represents a magnetic field $B_z$ splitting of the electronic spin $\left| \pm 1 \right>_{m_{s}}$ (nuclear spin $\left| \pm \frac{1}{2} \right>_{m_{n}}$) states with the Bohr magneton (nuclear magneton) given by ${\mu_{B}}/{2\pi\hbar}=14$ GHzT$^{-1}$ (${\mu_{N}}/{2\pi\hbar}=7.63$ MHzT$^{-1}$). The g-factors are $g^{||}_{GSM}=2.01$ and $g^{||}_{n}=-0.566$ respectively.

The structure of the excited state manifold is determined by the component $H_{ESM}$, where
\begin{eqnarray}
	H_{ESM} &=& \hbar D^{||}_{ESM} \left( S^{2}_{z} - \frac{2}{3} \right)L^{2}_{z} - \hbar \lambda^{||}_{ESM} S_{z} L_{z} \nonumber \\
	&+&\frac{\hbar}{2} D^{\perp}_{ESM} \left( S^{2}_{y} - S^{2}_{x} \right) \left( L^{2}_{x} - L^{2}_{y} \right)\nonumber \\
	&-& \frac{\hbar}{2} D^{\perp}_{ESM} \left( S_{y}S_{x} + S_{x}S_{y} \right) \left( L_{x}L_{y} + L_{y}L_{x} \right)\nonumber\\
	&+&\frac{\hbar}{2} \lambda^{\perp}_{ESM}\left( S_{x}S_{z} + S_{z}S_{x} \right) \left( L^{2}_{x} - L^{2}_{y} \right) \nonumber \\
	&-&\frac{\hbar}{2} \lambda^{\perp}_{ESM}\left( S_{y}S_{z} + S_{z}S_{y} \right) \left( L_{x}L_{y} + L_{y}L_{x} \right)  \nonumber \\
        &+& \mu_{B} \left( l^{||}_{ESM} L_{z} + g^{||}_{ESM} S_{z}L^{2}_{z} \right) B_{z}.
	\label{eqn:HamESM}
\end{eqnarray}
Here ${D^{||}_{ESM}}/{2\pi}=1.21$ GHz denotes the zero-field splitting, ${D^{\perp}_{ESM}}/{2\pi}=0.6375$ GHz, ${\lambda^{||}_{ESM}}/{2\pi}=4.85$ GHz, and ${\lambda^{\perp}_{ESM}}/{2\pi}=0.141$ GHz are spin-orbit interaction terms and $g^{||}_{ESM},g^{\perp}_{ESM}=2.01$ ($l^{||}_{ESM}=0.1$) are the electronic spin (orbital angular momentum) g-factors respectively. At cryogenic temperatures ($4-8$ K) the non-zero orbital angular momentum components are distinct \cite{Tamarat06,Batalov09} and determine photon polarization selection rules on allowed optical transitions \cite{Chu15}. These are enumerated in Table \ref{tbl:zeroFieldES} (along with free transition lifetimes \cite{Goldman15,Collins83}) for the zero-field and 20 mT external magnetic field cases \footnote{We have assumed ideal control over the electric and magnetic fields; we expect, though the effects are not investigated here, that a transition to the large magnetic field regime would suppress variation in the electric field (or due to strain). As noted by Childress et al. \cite{Childress05}, for weak driving, fluctuations in the transition energies over time have the effect of broadening the transition, combining with the decay rate to reduce the cooperativity.}. The zero-phonon line between the $\left|0\right>_{m_{s}}$ intrinsic spin ground state and the optically excited states at zero field is $637$nm. As the nitrogen--vacancy center decays,  emission into the phonon side-band collectively exceeds emission at the zero-phonon line  \cite{Davies74,Davies76,Kehayias13}. Elastic scattering occurs approximately $3$--$5$\% of the time \cite{Jelezko02,Santori10,Albrecht14,Johnson15}, linearly degrading the cooperativity of a coupled cavity.

Next (as seen in Figure \ref{fig:full_structure}), a decay channel couples the optically excited states to a meta-stable subspace. These decay rates (shown in Table \ref{tbl:zeroFieldES}) can have a significant effect on the measurement fidelity as decay through a meta-stable state removes phase information and introduce a bit-flip error rate of $67$--$81$\% from the $m_{s}=+1$ state, and $38$--$65$\% from the $m_{s}=0$ state \cite{Robledo11b,Tetienne12}. Here we assume a polarizing sample with a bit-flip rate $81$\% from the $m_{s}=+1$ state, and $38$\% from the $m_{s}=0$ state. The longer-lived $^{1}$E meta-stable state has a lifetime of $462$ ns at $4.4$ K, while the shorter-lived $^{1}$A$_{1}$ state (separated by a $1042$ nm energy gap from its $^{1}$E counterpart)  has a lifetime of less than $1$ ns \cite{Acosta10}.  Transitions to and from the meta-stable subspace do not conserve the intrinsic electron spin \cite{Goldman15}, they therefore degrade the measurement fidelity.

Having now outlined the properties of the NV center essential to the measurement process, let us move to a description of our photonic readout.

\section{Photonic Readout of the Nitrogen--Vacancy Center State} 
\label{sec:photonic_readout_of_the_nitrogen_vacancy_center_state}

\begin{figure}[hbt]
\includegraphics[width=0.4\textwidth]{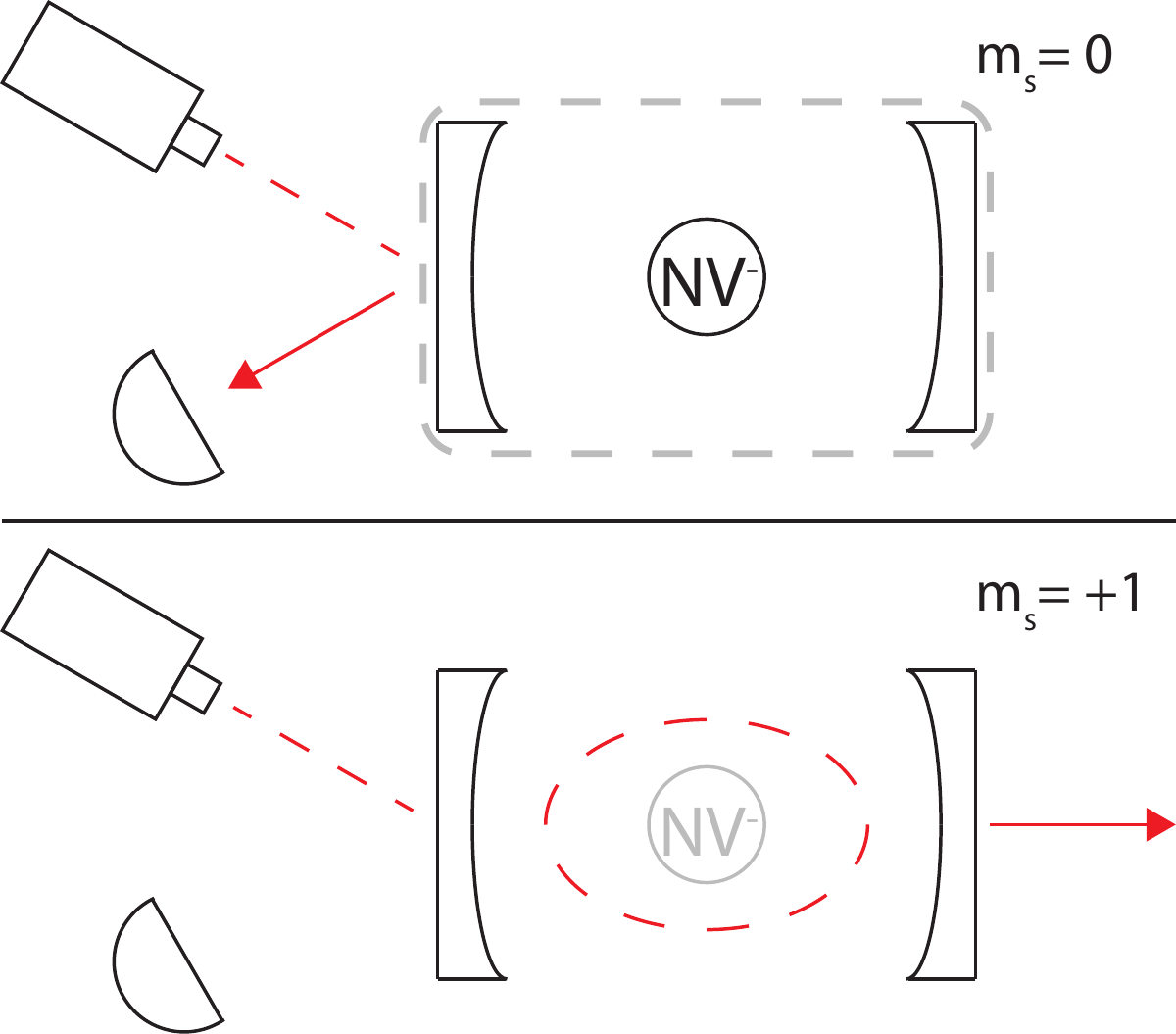}
\caption{When the nitrogen--vacancy center is in the $m_{s}=0$ ground state, resonant coupling with the optical cavity causes the composite system to form dressed states detuned from the frequency of the incident photon --- the photon does not enter the cavity, and is reflected. When the nitrogen--vacancy center is in the $m_{s}=+1$ state, there is no optical transition close enough in energy to be able to couple with the cavity mode --- the photon \emph{sees} an empty cavity, enters, and is transmitted. In such ideal cases, no interaction occurs between the incident photon and the nitrogen--vacancy center, preserving the center's state.}
\label{fig:two_step_measurement} 
\end{figure}

The optically accessible transitions in an NV center provide a natural way to measure it, given the excellent frequency separation of allowed transitions from the $m_s=0$ state relative to those of the $m_s=\pm 1$ states (Table \ref{tbl:zeroFieldES}).  The absence of hyperfine interactions \footnote{The hyperfine interaction is effectively turned off for the $m_s=0$ levels, while there is a difference in hyperfine coupling constants between the ESM and GSM $m_s=\pm 1$ levels. The spontaneous nature of the decay from the ESM will lead to a small random phase shift in this case.} and the smaller decay rates from the $m_s=0$ ESM levels to the meta-stable subspace leads to the selection of the $|0\rangle_{m_s} \leftrightarrow |M_{3,4}\rangle_{m_s}$ transitions for the optical readout. The presence of decay through a meta-stable state, however, limits that measurement efficiency after a certain number of light pulses, as the excited states $|M_{3,4}\rangle_{m_s}$ will decay into the MSM ($<1$\% of the time). Dipole-induced transparency (giant Faraday rotation) \cite{Nemoto14,Hu08,Young09,Koshino12,Sun16} provides an elegant way to mitigate this effect, as we can use strong coupling between our NV center and an optical cavity to modify the resonance properties of that cavity. In this strong coupling regime, a $\sigma_+$ polarized photon near resonance with the empty cavity and with the $|0\rangle_{m_s} \leftrightarrow |M_{4}\rangle_{m_s}$ transition would be reflected when the NV center is in the $|0\rangle_{m_s}$ state \footnote{The use of a $\sigma_+$ polarized photon means that photon can not be resonant with the $\left|0\right>_{m_{s}}$ -- $\left|M3\right>$ transition.}. 
This means the photon is not absorbed by the NV center, mitigating the effect of decay through the meta-stable state and providing, in the ideal case, an interaction-free measurement. On the other hand, the photon will be transmitted through the cavity (and not absorbed or scattered) if the NV center is in its $|\pm 1 \rangle_{m_s}$ states as it is far off resonance with the  $|\pm 1\rangle_{m_s} \leftrightarrow |M_{1,2,5,6}\rangle_{m_s}$ transitions. Measurement of the presence or absence of the reflected photon (or series of photons) thus allows us to infer the state of the NV's electron spin.  How well this works required a detailed analysis of  the entire measurement scheme.

Coupling between the cavity, external field modes and electron spin can be represented as shown in Figure \ref{fig:full_structure} and by the Hamiltonian
\begin{eqnarray}
	H_{\rm coupling} &= & \hbar  \left[ c \left(\sqrt{\frac{\kappa_{a}}{\pi}} a^{\dagger}  + \sqrt{\frac{\kappa_{b}}{\pi}} b^{\dagger} \right) + \sqrt{2} g  c L_x \right]  \nonumber \\
	&+& \hbar \sum^{6}_{i=1} \sqrt{\frac{2\gamma_{i}}{\pi}}  d  |M_i\rangle \langle M_i | L_x  +{\rm h.c},
\label{eqn:HamCoupling}
\end{eqnarray}
where $a,b,c,d$ ($a^\dagger,b^\dagger,c^\dagger,d^\dagger$) are the annihilation (creation) operators of the left-hand, right-hand, cavity and scattering (spontaneous emission) operators. Next, $L_x$ is the angular spin-1 X operator (see Appendix \ref{sec:spin_operators}) while $g$ is the vacuum-Rabi coupling rate. Further, $\kappa_{a}$ and $\kappa_{b}$ are the left and right mirror cavity decay rates, which we assume are equal giving us a total decay rate $\kappa=\kappa_{a}+\kappa_{b}$. Finally, $\gamma_{i}$ is half the spontaneous decay rate of the $i$th optically excited state \footnote{These have not be measured at 20 mT but their values can be estimated roughly by appropriate mixing of the zero-field values.}. These three parameters allow us to define the cooperativity $C_i= g^{2}/2\kappa\gamma_i$, a useful measure for determining how strongly coupled one is \cite{Bonifacio76}.

On resonance, the coupling is related to the spontaneous decay rate by
\begin{eqnarray}
	g &= \sqrt{\rho_{\omega}\frac{\pi^{2}c^{3}}{\hbar \omega^{3}}\frac{\gamma_{Rad}}{2\pi} f},
\end{eqnarray}
where $f$ is the quantum efficiency of the transition ($0.03-0.05$ for the nitrogen--vacancy center), $\gamma_{Rad}$ is the radiative component of the spontaneous decay rate ($2\pi \times 13.2$ MHz), $\omega$ is the angular frequency of the transition, $c$ is the speed of light, and $\rho_{\omega}$ denotes the energy density of the cavity-field per unit angular frequency \cite{Hilborn82}.

With the NV and coupling Hamiltonians we can derive Langevin equations of motion for the various field and spin operators. These are nonlinear in nature due to coupling terms between the field and spin operators, which makes them difficult to solve in general. Two alternative approaches can be taken to simplify this situation: the first assumes that NV center is always in the GSM while the second is to work in a single excitation subspace \cite{Rephaeli12} where the field probing the NV center contains no more than one photon (both approaches lead to the same answer). We take the second route and will use probe fields with at most one photon. This approximation results in a set of linear Langevin equations that are straightforward but tedious to solve. The solutions are those for the cavity mode and excited states and not what we measure. However by using input--output relations \cite{Gardiner85}
\begin{align}
	a_{\text{out}}(t) = a_{\text{in}}(t) - i \sqrt{\kappa} c(t),
\end{align}
we can express our output field $a_{\text{out}}(t)$ in terms of the cavity field $c(t)$ and the cavity input field $a_{\text{in}}(t)$. This then enables us to calculate the mean photon number $\langle a^\dagger _{\text{out}}(t) a_{\text{out}}(t) \rangle$ over time and so determine whether an incident photon was reflected or not.

Source and detection efficiencies, as well as losses and imperfections, mean that one single-photon measurement will not be sufficient to determine the NV state with the accuracy we require. Instead our measurement here consists of a series of temporally spaced single-photon pulses which we individually attempt to detect. The time between photon pulses is chosen to be $\sim165$ ns, from the axial hyperfine Rabi period. By matching gate times to this period, the effect of electron dephasing on the nuclear state can be minimized. Though it is not possible to restrict pulse times exactly to a periodic point, it has been shown that error associated with decay channels following optical excitation can be reduced by centering the pulse on such a point \cite{Reiserer16}.  This is much larger than the excited state lifetime of the $|M_{4}\rangle$ state ($12.1$ ns) \cite{Goldman15}, but significantly shorter than the GSM decoherence times.

The outcome at the detector of each measurement pulse is dependent on its initial state distribution and thus successive outcomes are dependent on the detection history preceding them. After $n$ pulses therefore there are $2^{n}$ possible measurement paths. Not all branches in the outcome tree will be full-length, however, as outcome branches that are successful early in the procedure can be truncated. In particular, reflection from the $\left| +1 \right>_{m_{s}}$ state can be suppressed by polarization and detuning, so that a detection event very strongly indicates the $|0\rangle_{m_{s}}$ state and will terminate an outcome branch (especially as dark count probabilities can be exceptionally low \cite{FirstSensor16,ExcelitasTechnologies16,LaserComponents16}).

\section{Simulation of the Measurement Process} 
\label{sec:analysis}

\begin{figure}[hbt]
\includegraphics[width=0.48\textwidth]{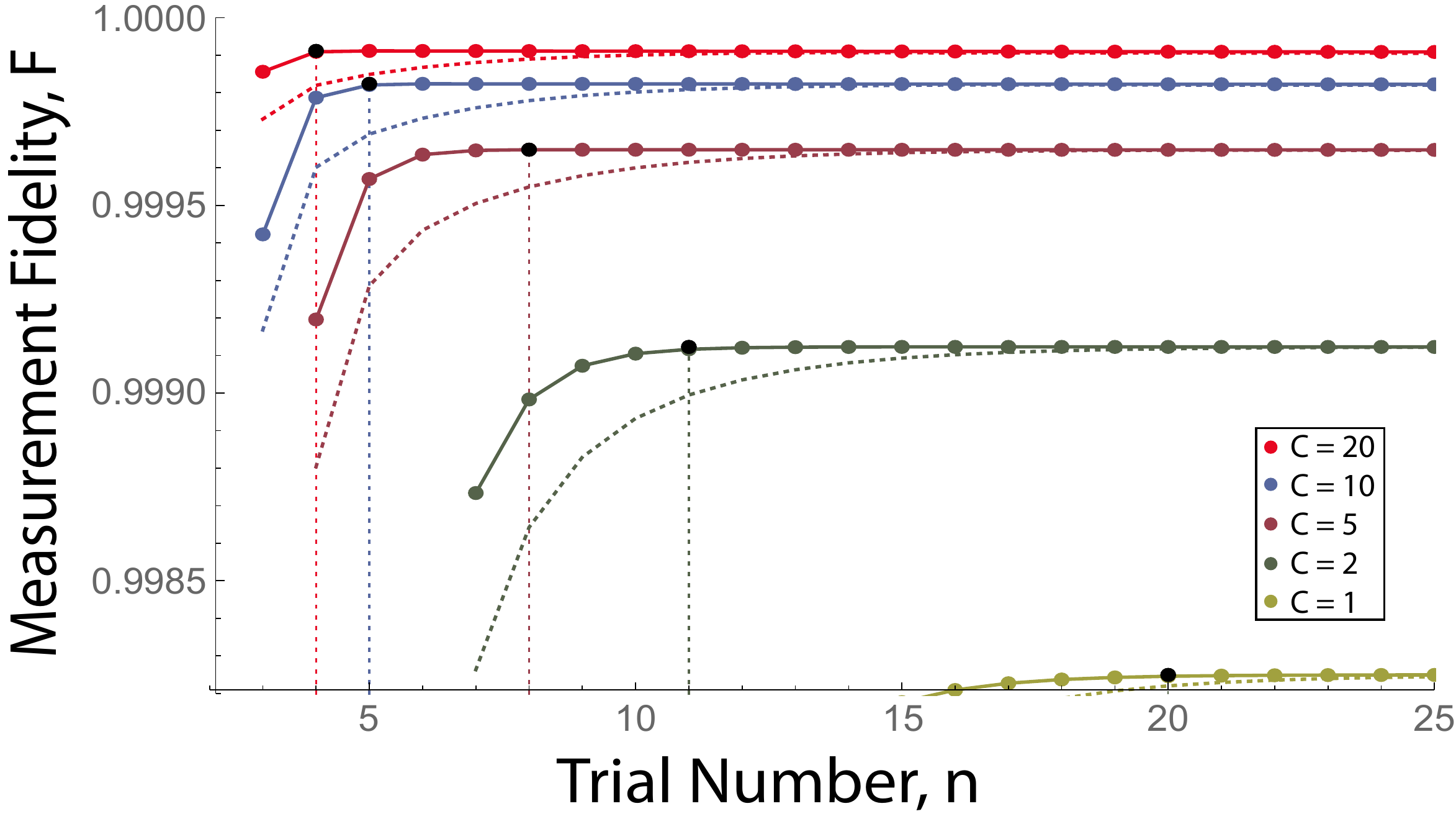}
\includegraphics[width=0.48\textwidth]{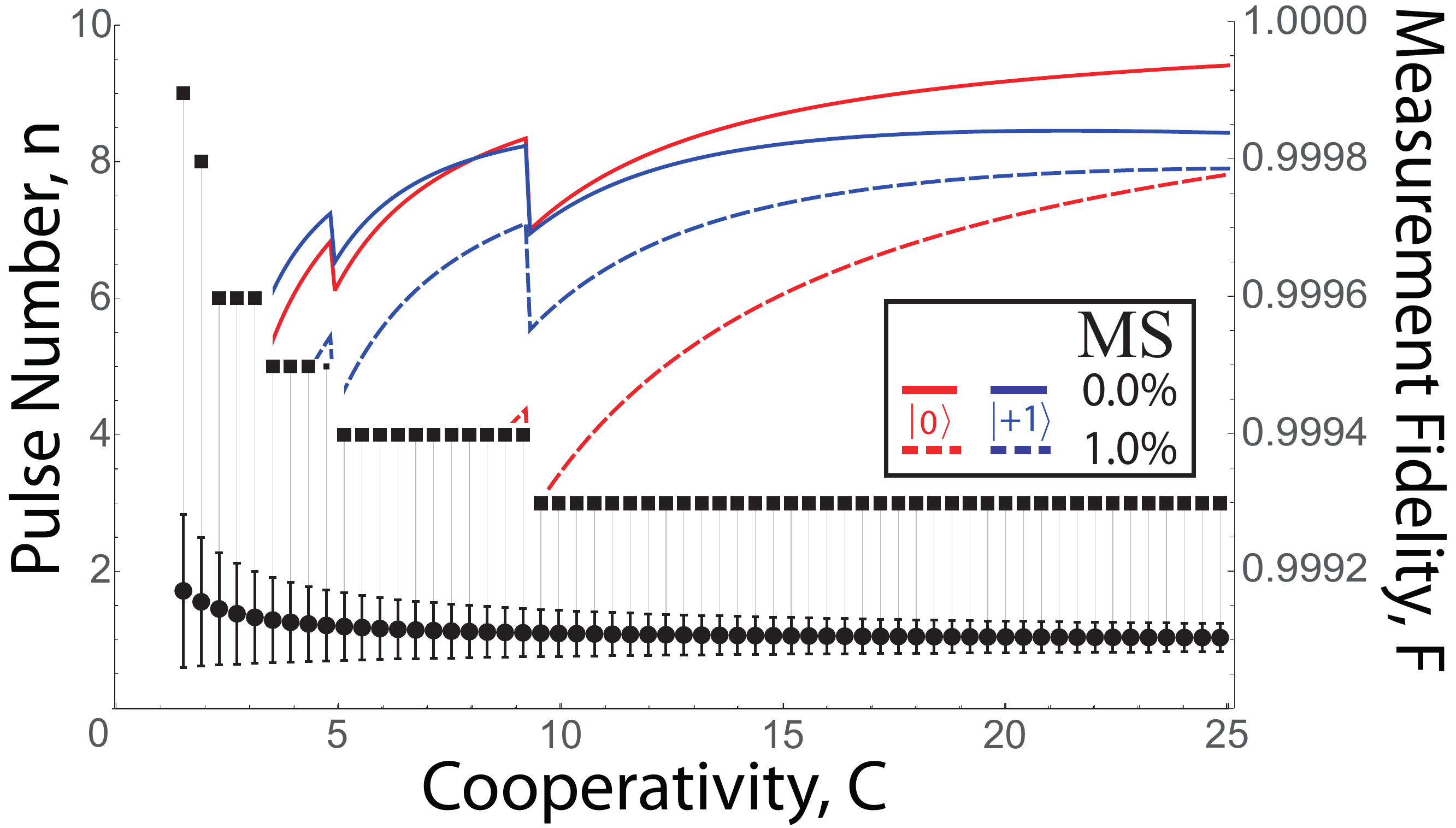}
\caption{
{\bf (Top)} The measurement fidelity as a function of the number of measurement trials performed, displayed for a set of cooperativity values from $1$ to $20$. Peak numbers are indicated by black dots and with dashed lines. Dotted lines indicate the performance when the $m_{s}=0$ decay to the metastable subspace is $1$\%.
{\bf (Bottom, Left)} The number of trials $n$ to obtain a positive measurement result (a reflected photon or `click') for an electronic spin prepared in the $\left| 0 \right>_{m_{s}}$ state, as a function of the cooperativity. Depicted are the mean number $n_{ave}$ (solid circles) with one standard deviation error bars and $n_{ft}$, the number of trials needed for a single photon to be reflected with probability $p_{s} \geq 0.999$ (solid squares), rounded up to the nearest integer value. Numbers are calculated based on the reflection probability of an initial pure state.
{\bf (Bottom, Right)} The measurement fidelities $F_{0}$ (red curve) and $F_{+1}$ (blue curve) for varying $m_{s}=0$ transition rates through the meta-stable subspace (expressed as a percentage of the total spontaneous decay rate).
 The discrete drops in the measurement fidelities are due to the integer nature of the number of trials required which drops as the cooperativity increases.}
\label{fig:ideal_fidelity} 
\end{figure}

\begin{figure}[htb]
\includegraphics[width=0.45\textwidth]{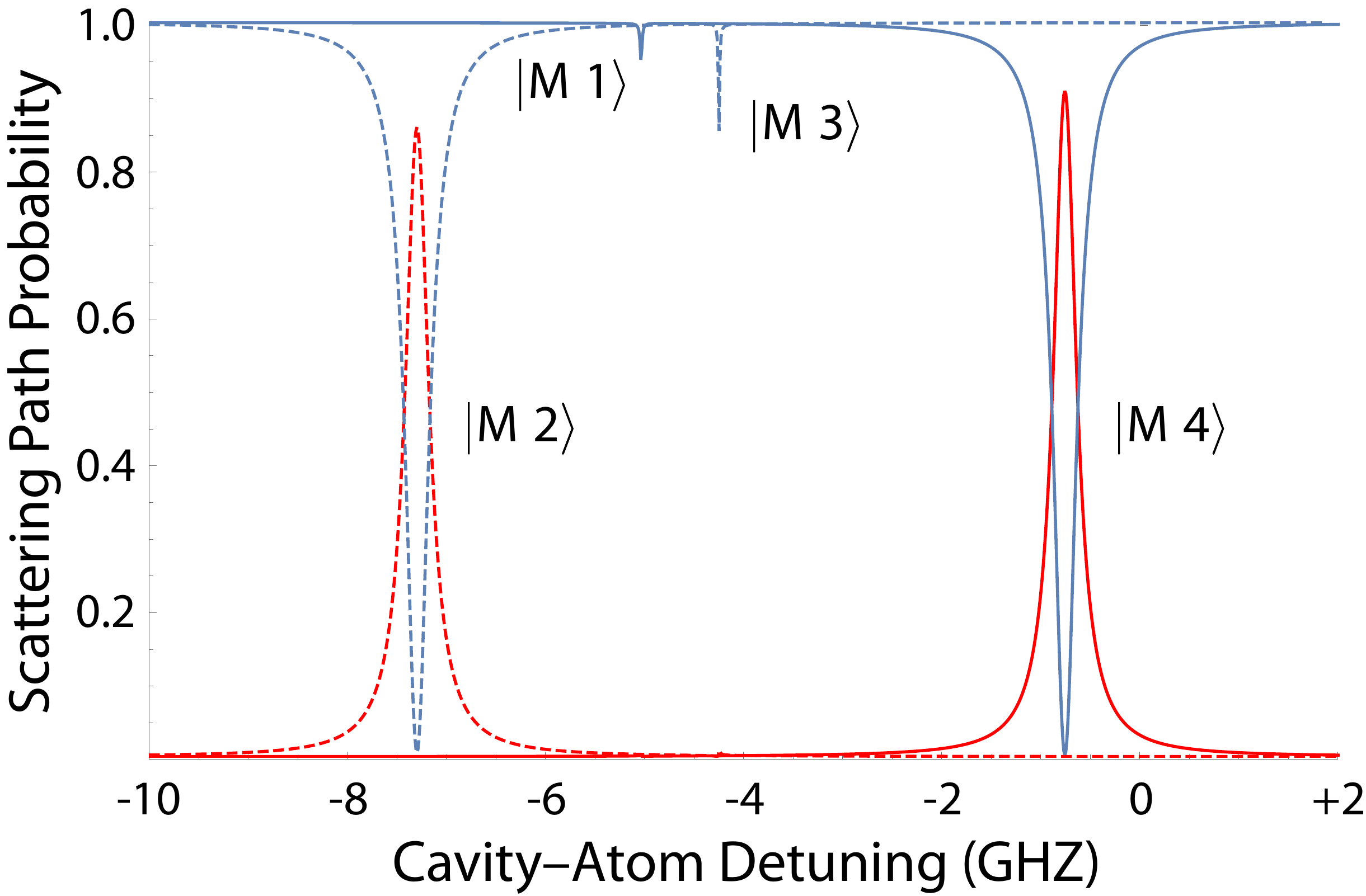}
\caption{Reflection (red) and transmission (blue) spectra, over a range of cavity-atom detunings relative to the energies depicted in Figure \ref{fig:full_structure}, when the nitrogen--vacancy center is in the $\left| 0 \right>_{m_{s}}$ (solid line) and $\left| +1 \right>_{m_{s}}$ (dashed line) states. $\sigma_{+}$ photon polarization has removed several other possible transitions. The cooperativity is taken at $C=10$. The presence of additional decay paths prevents these lines from summing to one.}
\label{fig_spectra} 
\end{figure}

Along with our description of the NV$^-$ center's photonic readout, let us now turn our attention to simulating it. Our model has a large number of parameters, but with the external magnetic (electric) fields set at 20 mT (0 V/m), our primary focus will be restricted to four experimentally relevant parameters: the cooperativity $C$, single photon source and detection efficiencies and the $m_{s}=0$ decay rates through the meta-stable subspace. Figure \ref{fig_spectra} shows example transmission and reflection spectra.

It is useful to begin by considering the situation with perfect single photon sources and detectors but with finite cooperativity $C$ and transitions through the meta-stable subspace (in the range $0$--$1$\% for the target transition, as indicated in Table \ref{tbl:zeroFieldES}). When the electronic spin is in its $\left| 0 \right>_{m_{s}}$ state, an incoming photon is reflected with a probability {$P_{R} \approx 4 C^2 / (2 C+1)^{2}$}, so that if we send $n$ photons, the probability of detecting one or more photons is $p_{s} = 1-\left(1-P_R\right)^n$ (equivalently we need $n = \log (1-p_{s}) / \log (1-P_{R})$ trials to have at least one photon detected with success probability $p_{s}$). There are two $n$'s of interest here,  $n_{ave}$ and $n_{ft}$. $n_{ave}$ is associated with the average number of single photon trials to get a measurement `click' while  $n_{ft}$ is the number that guarantees at least one measurement click with probability above the fault tolerance threshold we set of $99.9$\%. For an initial state $\left| +1 \right>_{m_{s}}$ we have to perform $n_{ft}$ trials (we call this trial limit number the {\it stop-limit}) with no clicks to infer it was this state (with error probability $1- p_{s}$). Figure \ref{fig:ideal_fidelity} shows the effect of the cooperativity on both $n_{ave}, n_{ft}$ and the measurement fidelities $F_{i \in \{0,+1\}}=\langle i| \rho | i \rangle_{m_{s}}$ (where $\rho$ is the state of the system after the $n$ trials) for varying decay rates to the meta-stable subspace. As the cooperativity, and therefore also the reflection probability, increases, the number of trials required to achieve $p_{s}$ decreases in integer steps. These steps manifest as negative discontinuous jumps in the measurement fidelity.

\begin{figure}[b]
\includegraphics[width=0.49\textwidth]{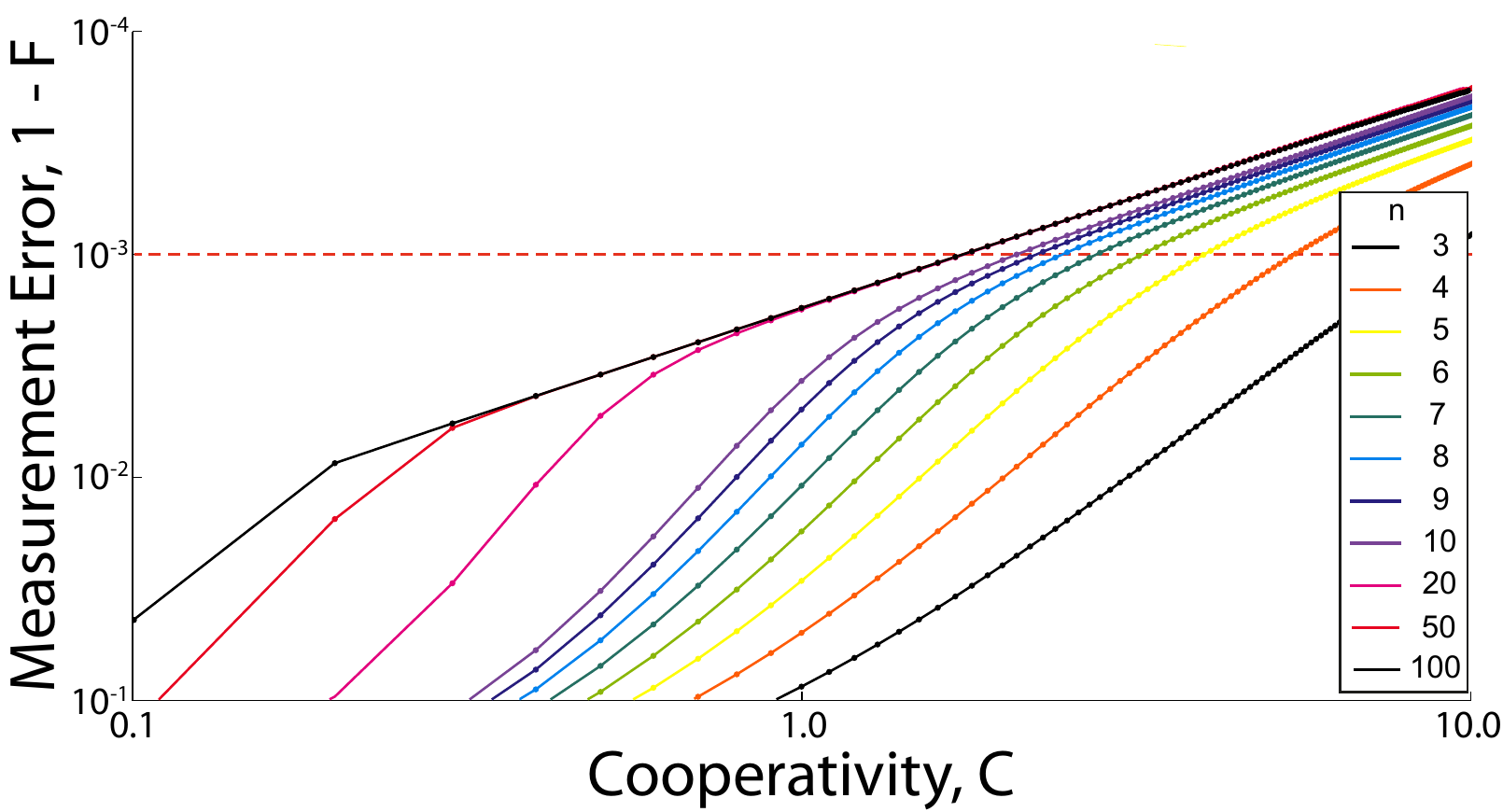}
\caption{\label{fig:ideal_performance} Loss of measurement fidelity, $F$, as function of the cooperativity, $C$, for an ideal single-photon source and detector, over a range of measurement pulses, $n$. The $m_{s}=0$ decay rate to the meta-stable subspace is $1$\%. While the optimal number of pulses decreases with increasing cooperativity, the error associated with moving away from this optimum is very small ($\sim 0.0001$) for cooperativities above $5$.}
\end{figure}

When we start with a general superposition state, each measurement pulse will introduce some error and so the reflection probability will not remain static as was assumed in Figure \ref{fig:ideal_fidelity}. Instead, the contrast between the initial $\left| 0 \right>_{m_{s}}$ and $\left| +1 \right>_{m_{s}}$ states will degrade as the number of pulses increases, thereby increasing $n_{ft}$.
Rather than initially specifying a maximum number of trials, we can look at the performance over a range of such numbers to identify an optimum number of measurement trials and the sensitivity of the measurement performance to changes in this number.
In Figure \ref{fig:ideal_performance} we plot the measurement fidelity for the initial state $\left( \left| 0 \right>_{m_{s}} \left| -\frac{1}{2} \right>_{m_{n}} + \left| +1 \right>_{m_{s}} \left| +\frac{1}{2} \right>_{m_{n}} \right) / \sqrt{2}$. For the ideal case we observe that the $99.9$\% threshold is met even at cooperativities as low as $C = 2$, and that the dependence of the measurement fidelity on the pulse number, while sensitive at low cooperativities, is as low as order $0.0001$ for cooperativities greater than $5$.

\begin{figure}[tb]
\includegraphics[width=0.42\textwidth]{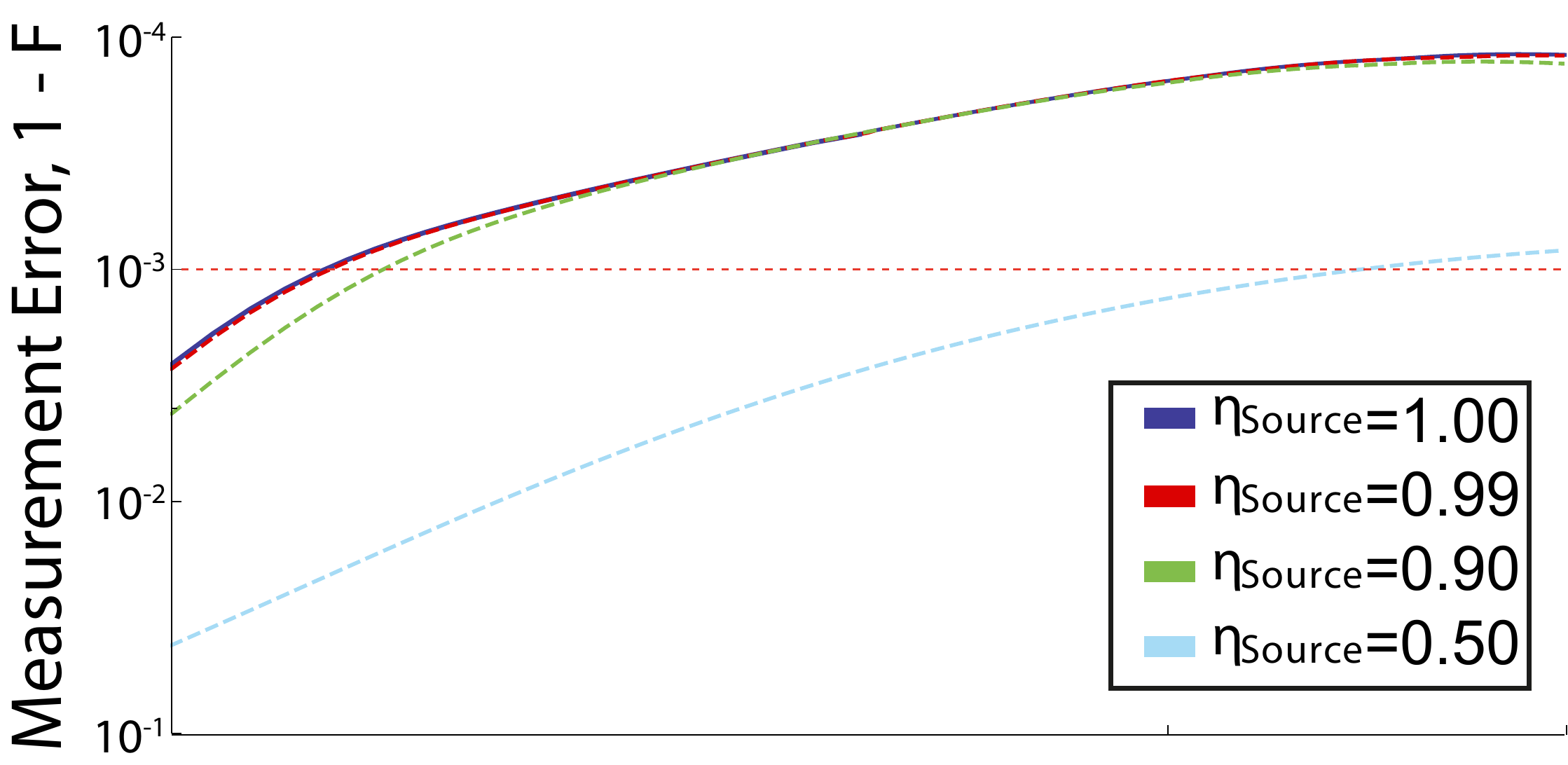}
\includegraphics[width=0.42\textwidth]{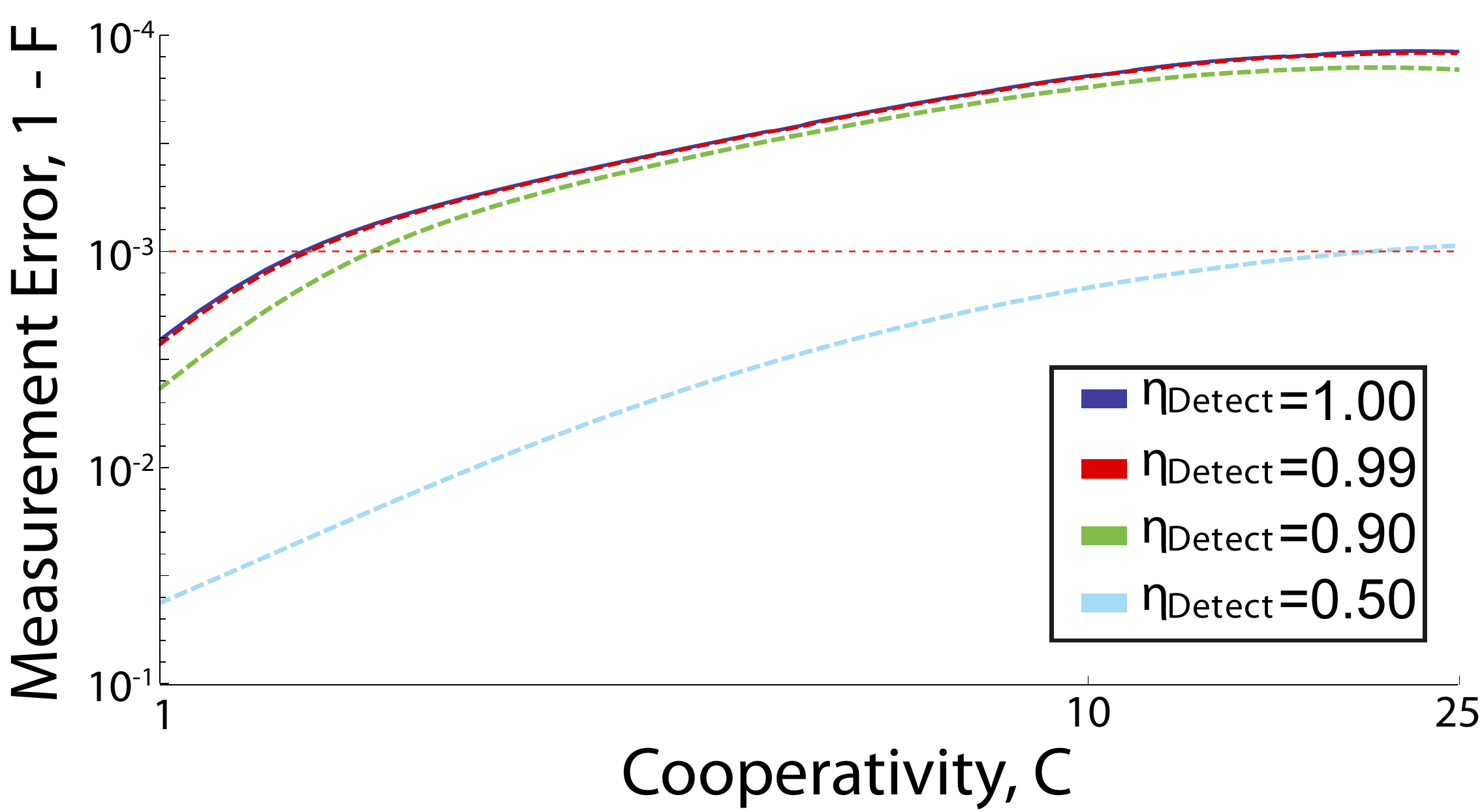}
\caption{\label{fig:external_parameters} Loss of measurement fidelity, $F$, for a maximum of 10 photon pulses and a system initialized in the electronic and nuclear spin state $\left( \left| 0 \right>_{m_{s}} \left| -\frac{1}{2} \right>_{m_{n}}  + \left| +1 \right>_{m_{s}} \left| +\frac{1}{2} \right>_{m_{n}} \right) / \sqrt{2}$, as a function of the cooperativity, $C$, and {\bf (Top)} single-photon source efficiency, $\eta_{Source}$, and {\bf (Bottom)} single-photon detection efficiency, $\eta_{Detect}$. The $m_{s}=0$ decay rate to the meta-stable subspace is taken to be $1$\%.}
\end{figure}

Now including the effects of source and detection efficiency into our analysis, we show in Figure \ref{fig:external_parameters} the behavior of the measurement fidelity as a function of source and detection efficiencies. We observe that the performance depends slightly more sensitively on the detector than the source, but that the performance is robust for small ($\sim 0.1$) inefficiencies.

\subsection{Realistic Performance Estimate} 
\label{sec:performance_estimate_with_commercial_parameters}

A natural question to ask is what our expected measurement fidelity would be with current technology. To this end let us specify a cooperativity of $0.2$ \cite{Janitz15}, a single-photon source probability of $60$\% \footnote{PPLN sources can be purchased with $60$\% heralding efficiency \cite{Idquantique16}, which may be multiplexed at a low individual extraction rate to suppress multi-photon components. Recent genuine single-photon sources have also been built based on quantum dots \cite{Somaschi16} that have achieved extraction efficiencies of $65$\%, though unpolarized and at a different frequency than the one we require}, a single-photon detection efficiency of $92$\%  \cite{QuantumOpus16,PhotonSpot16,Scontel16,SingleQuantum16,ExcelitasTechnologies16,LaserComponents16}, and $m_{s}=0$ decay rates to the meta-stable subspace of $1$\% \cite{Goldman15}. Our model then predicts a maximum measurement fidelity of $F=0.992$ occurring (within rounding error) after $145$ single photon pulses. The total time for the measurement is then $\sim 150\times165$ ns $\approx 25\:\mu$s. During this time we would expect an electron spin dephasing error between $1$\%--$0.01$\% for coherence times in the region $1$--$100$ ms. The upper range would have a consequential effect on remote entangling operations, though dephasing should not effect the nuclear spin measurements. Limiting the maximum number of pulses to $10$ reduces the measurement fidelity to $F=0.686$. In Table \ref{tbl:realParamPerformance} for contrast we outline similar numbers for $C\in\{0.5,1,2,5,10\}$ and $\eta_{Source},\eta_{Detect}\in\{0.2,0.6\}$. For a low source efficiency, it is apparent that high measurement fidelities can be maintained at the cost of greatly increasing the number of pulses.

\begin{table}[tbht]
\caption{\label{tbl:realParamPerformance} Measurement fidelities, $F$, and pulse numbers, $n$, when the $m_{s}=0$ decay rate to the meta-stable subspace is $1$\%, varying the cooperativity and the single-photon source and detection efficiencies, $\eta_{\text{Source}}$ and $\eta_{\text{Detect}}$.}
\begin{ruledtabular}
\begin{tabular}{ c | c | c | c | c }
		C & $\eta_{\text{Source}}$ & $\eta_{\text{Detect}}$ & F & n \\
		\hline
		 $0.5$ & $0.2$ & $1.0$ & $0.9965$ & $290$ \\
		 $1$ & $0.2$ & $1.0$ & $0.9982$ & $159$ \\
		 $2$ & $0.2$ & $1.0$ & $0.9992$ & $80$ \\
		 $5$ & $0.2$ & $1.0$ & $0.9997$ & $66$ \\
		$10$ & $0.2$ & $1.0$ & $0.9998$ & $57$ \\
		\hline
		 $0.5$ & $0.6$ & $1.0$ & $0.9965$ & $93$ \\
		 $1$ & $0.6$ & $1.0$ & $0.9985$ & $42$ \\
		 $2$ & $0.6$ & $1.0$ & $0.9992$ & $27$ \\
		 $5$ & $0.6$ & $1.0$ & $0.9997$ & $19$ \\
		$10$ & $0.6$ & $1.0$ & $0.9998$ & $15$\\
		\hline
		 $0.5$ & $1.0$ & $0.2$ & $0.9830$ & $255$ \\
		 $1$ & $1.0$ & $0.2$ & $0.9914$ & $141$ \\
		 $2$ & $1.0$ & $0.2$ & $0.9962$ & $73$ \\
		 $5$ & $1.0$ & $0.2$ & $0.9983$ & $57$ \\
		$10$ & $1.0$ & $0.2$ & $0.9990$ & $49$ \\
		\hline
		 $0.5$ & $1.0$ & $0.6$ & $0.9950$ & $72$ \\
		 $1$ & $1.0$ & $0.6$ & $0.9975$ & $41$ \\
		 $2$ & $1.0$ & $0.6$ & $0.9987$ & $24$ \\
		 $5$ & $1.0$ & $0.6$ & $0.9995$ & $18$ \\
		$10$ & $1.0$ & $0.6$ & $0.9997$ & $15$
\end{tabular}
\end{ruledtabular}
\end{table}

\subsection{Single Photon Pulses} 
\label{sec:finite_bandwidth_photons}

In any realistic model we also need to consider the bandwidth of our atomic resonances compared to those of the photon. Here we are working in the regime where the photon bandwidth is much less than the bandwidth of the NV center's optical transitions.  Our single photon pulses are $165$ ns apart implying a bandwidth at least of order $\Gamma_{P}/ 2 \pi \sim 1/2$ MHz (HWHM). The reflection spectrum from $\left|0\right>_{m_{s}}$ is not sensitive at those scales but reflection from $\left|+1\right>_{m_{s}}$ sits in the trough of a narrow resonance ($\Gamma_{R} \sim \kappa = 2\pi \times 50$ MHz), necessitating further consideration. We assume a Gaussian pulse shape,
\begin{align}
	\frac{1}{\sqrt{2\sigma^{2}_{t}\pi}}e^{\frac{-(t-165/2)^{2}}{2\sigma^{2}_{t}}},
\end{align}
and set the standard deviation to $\sigma_{t}=165/6$ ns so that the width of the pulse is much less than the time between pulses (at three standard deviations the area truncated in the tails is $0.3$\%). By the time--bandwith product, $\sigma^{2}_{t} \sigma^{2}_{P}=1/4$, the standard deviation in the frequency then becomes $\sigma_{P}/2\pi = 2.9$ MHz. By integrating over the probability-weighted reflection spectrum we can estimate the realistic reflection probability to be
\begin{align}
	P_{ave\;R} &\sim 1 - \sqrt{\pi}\: (\sigma_{t}\kappa)\: \text{erfcx}\left(\sigma_{t}\kappa\right)\nonumber\\
	&= 0.66\%
\end{align}
The reflection here is a direct result of the relative magnitudes of the photon and cavity linewidths, and is not influenced by the nitrogen--vacancy center. Since the width of the reflection trough is set by $\kappa$, it also governs the transmission and scattering peaks, which are each reduced to $99.34$\% of their maximum values. A $0.66$\% reflection probability from the $\left|+1\right>_{m_{s}}$ state, when we want to suppress our error to within $0.1$\%, means that detection is no longer a strongly-classifying event.

Let us assume our initial state is predetermined and restricted to the \{$\left|0\right>_{m_{s}}$ and $\left|+1\right>_{m_{s}}$\} subspace, but that the fidelity decays exponentially in the number of photon pulses. The probability of each state after $n$ pulses is then given by Bayes' theorem and a binomial distribution (similar to the treatment of a Poisson-distribution in \cite{Sun16}). The probability of a detection event from $\left|0\right>_{m_{s}}$ and $\left|+1\right>_{m_{s}}$ will be $p_{0}$ and $p_{+1}$ respectively. With single-photon source efficiency, detection efficiency and cooperativity of $60$\%, $92$\% and $10$ respectively, the photon bandwidth detection probabilities are $p_{0}=0.50$ and $p_{+1}=0.00364$. We can estimate the error rates per pulse, as $\eta_{0}=1.5\times10^{-4}\; (\eta_{0}=3.5\times10^{-4}$) for the $m_{s}=0$ decay to the meta-stable subspace of $0$\% ($1$\%) respectively with $\eta_{+1}=1.2\times10^{-5}$. This means we can achieve an expected measurement fidelity of $F=99.9\%\; (F=99.8$\%) using $13\;(12$) single photon pulses. Using more pulses than this decreases the resulting fidelity but at the levels indicated here we are right at the border of our $99.9$\% requirement.

At what pulse times does the error associated with false-positive detection events become significant? We can obtain a quick estimate by determining the error and the average number of additional trials that would need to be performed to distinguish between the two rates of reflection, assuming once more that the error follows an exponential decay. As above, the cooperativity is assumed to be $10$, the source and detection probabilities are $60$\% and $92$\% respectively, and the $m_{s}=0$ metastable decay rate is $1$\%. In the point-frequency case this approximation predicts a fidelity of $0.99775$ after $11$ trials (this is a more conservative estimate than the numerical calculations represented in previous sections). Errors differing from this by $10^{-5}$, $10^{-4}$, and $10^{-3}$ occur when the pulse times reach $455$ ns (requiring $11$ trials), $115$ ns (requiring $12$ trials), and $24$ ns (requiring $19$ trials) respectively.

Understanding that the above is a conservative approximation, we can also ask what values for $\kappa$ or for $\sigma_{t}$ we would require to suppress reflection (rather than error) from the $\left|+1\right>_{m_{s}}$ so that it is below our threshold of $0.1$\%. We find that either $\kappa = 2\pi\times 129$ MHz for the current pulse-time or $\sigma_{t}=71$ ns for the current cavity decay rate (so that either is $2.58$ times larger than its original value) satisfy this condition. Due to dephasing, for application to projective entanglement generation there is a fundamental tradeoff to be made between longer pulse times, related directly to $\sigma_{t}$, and the number of such pulses required, as influenced by $\kappa$ through the cooperativity.

\section{Weak Coherent Pulses} 
\label{sec:wcp}

Our previous considerations (and in \cite{Nemoto14}) assumed the use of single-photon sources that are technological challenging to realize and were found to be a key limiting factor in achieving higher measurement fidelities. Weak coherent laser pulses are a natural alternative \cite{Volz11} and offer a number of potential advantages including:
\begin{itemize}
\item Ready availability at the appropriate wavelengths, 
\item Easy tailoring of their mean photon number,
\item Ready pulse shaping to customize the state dependent reflection/absorption from the cavity.
\end{itemize}
Allowing more than one photon to be reflected from the cavity dramatically improves our detection efficiency,  potentially turning this into a single-shot (pulse) measurement. There is however a potential issue here associated with ionization of the NV center, which can occur when the electronic state absorbs more than one 637 nm photon. Ionization converts the  NV$^-$ center to the charge neutral NV$^0$ center (which is spin-0). Little is known about photo-induced ionization at this wavelength and temperature apart from the fact it must be a two-photon process \cite{Beha12,Siyushev13,Aslam13}.  Once more is known , it may become possible to manipulate the exciting pulse to suppress them. 

\begin{figure}[htb]
\includegraphics[width=0.45\textwidth]{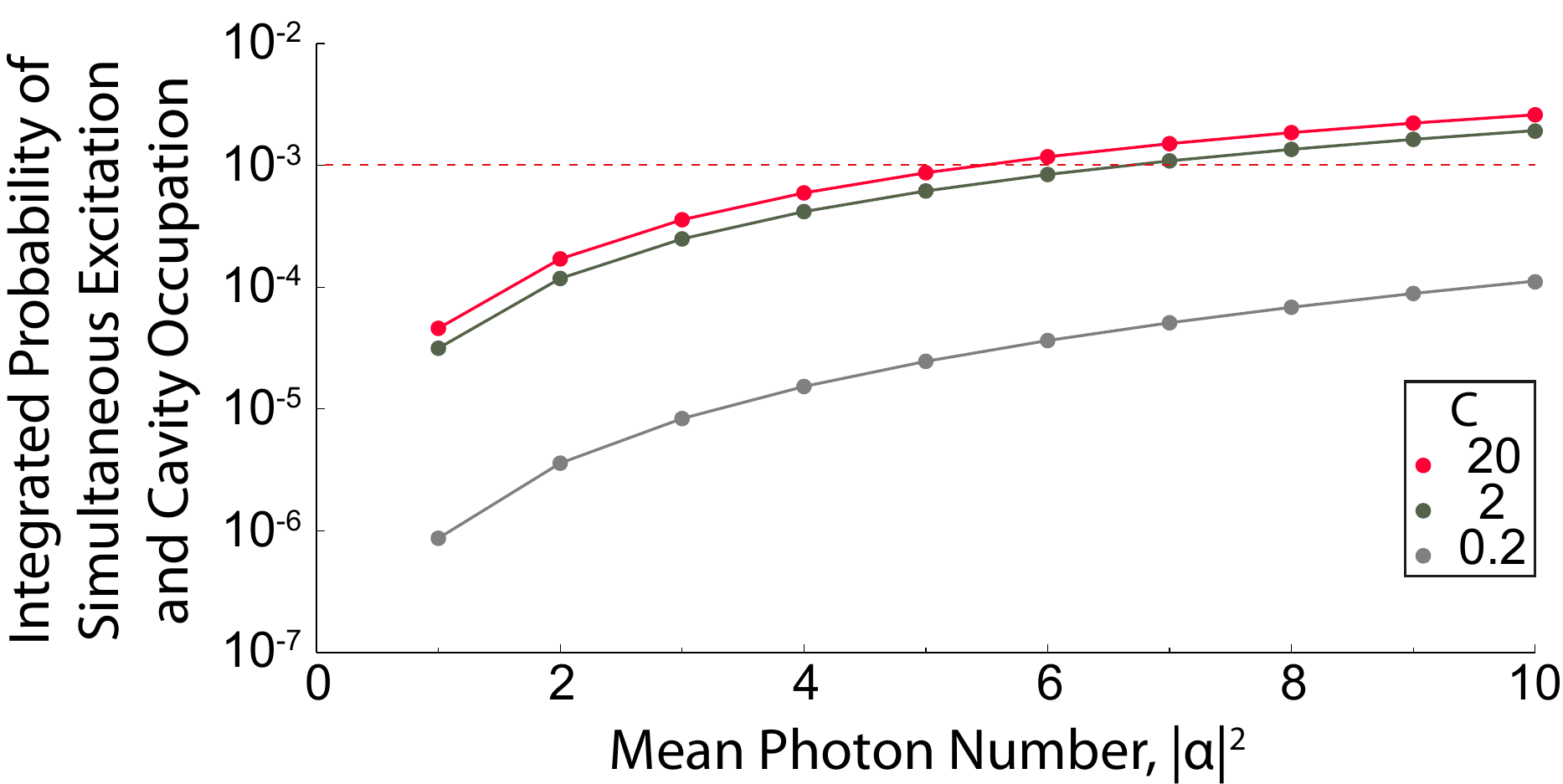}
\caption{Plot of the joint probability of the NV center being in the excited $|M_4\rangle$ state with one or more photons present in the cavity, integrated over the time of a single pulse, as a function of the mean photon number of that coherent pulse. This probability will serve as an upper bound on the ionization rate of the center. Shown for cooperativity values $C$ of $0.2$, $2$ and $20$.}
\label{fig:wcp}
\end{figure}

\begin{figure}[b]
\includegraphics[width=0.49\textwidth]{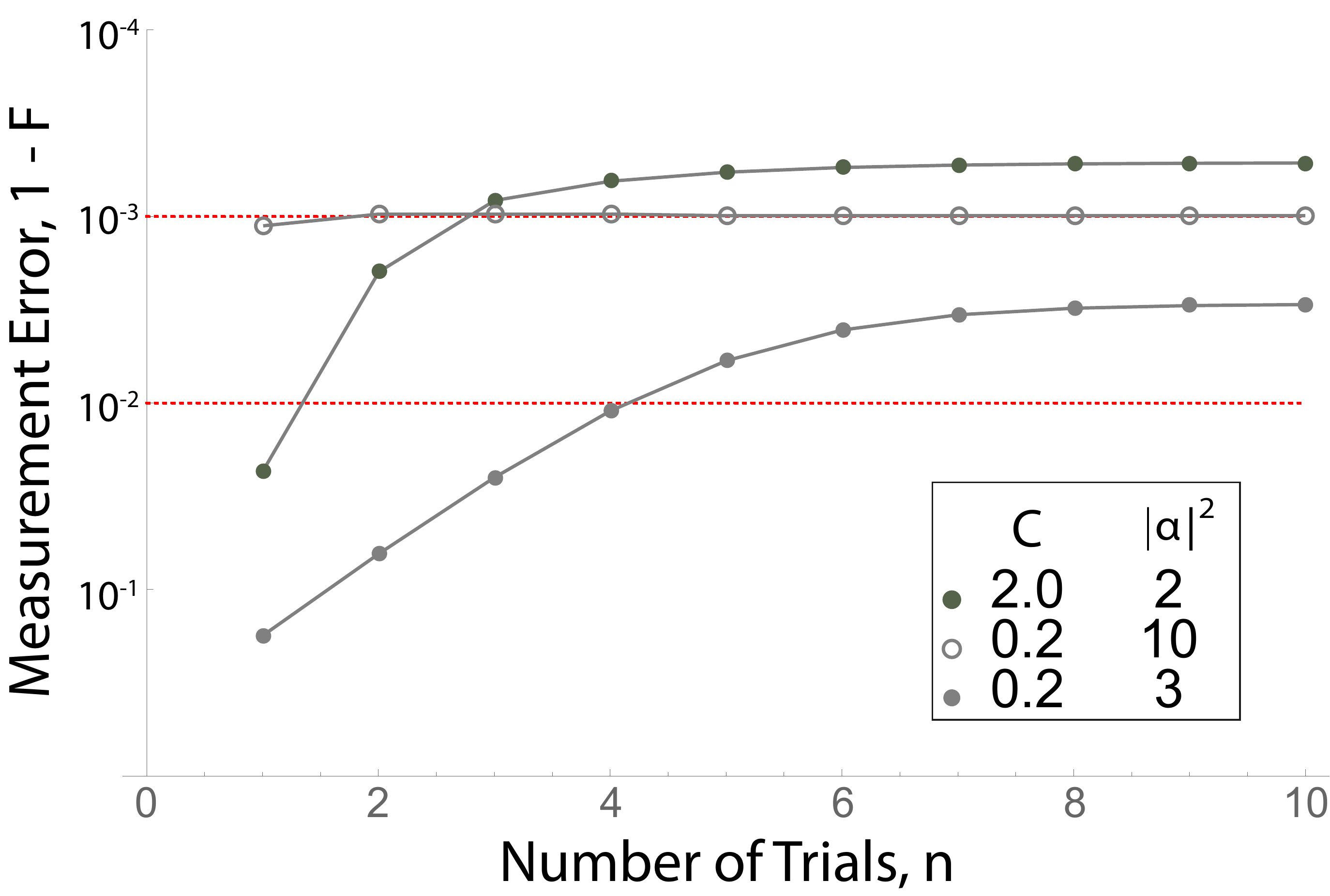}
\caption{\label{fig:coherent_ideal_performance} Loss of measurement fidelity, $F$, as function of the cooperativity, $C$, for a coherent light source and ideal detection, over a range of measurement pulses, $n$. The $m_{s}=0$ decay rate to the meta-stable subspace is taken to be $1$\%. Three cases are shown: $C=0.2,\left|\alpha\right|^{2}=3$ (light gray, dots), $C=0.2,\left|\alpha\right|^{2}=10$ (light gray, circles), $C=2.0,\left|\alpha\right|^{2}=2$ (dark gray, dots).}
\end{figure}

Our situation here is different as we are working with an NV in a CQED configuration at low temperatures, which will enhance the light--center interaction. We do however know that the probability of photons being reflected from the cavity mirror on resonance is {$P_{R} \approx 4 C^2/(2C+1)^2$} and for a weak coherent laser pulse $|\alpha\rangle$ the mean number of photons in that pulse entering the cavity is $|\alpha|^2 (1-P_{R})$. We can then estimate that over time of the entire pulse $\tau_{p}$, the probability $P_{2+}$ of having two or more photons within the cavity's line-width $1/\kappa$ should be 
\begin{eqnarray}
P_{2+}  \sim |\alpha|^4 (1-P_R)^2 \frac{\kappa \tau_p}{2}\sim 10^{-3}
\label{estimate}
\end{eqnarray}
for our typical parameters with $|\alpha|^2\sim 3$. It is critical to remember that having two photons in the cavity does not that mean both are simultaneously absorbed and cause ionization. We can go a step further and calculate the probability that the NV center is in its ESM and that simultaneously at least one photon is in the cavity mode. The results of our master equation simulation are shown in Figure \ref{fig:wcp} and we immediately observe that this joint probability is nearly an order of magnitude lower than $P_{2+}$. Unfortunately we do not know the coupling rate between the ESM and the conduction band. However, based on these preliminary estimates, for reasonable values of $C$ and $|\alpha|^2$ we expect that the ionization probability will be less than $10^{-3}$, and that therefore weak coherent laser pulses should be ideal for initial measurement experiments. Performance estimates for several cases are shown in Figure \ref{fig:coherent_ideal_performance}. In addition to increasing the  ionization  rate, a higher photon number also increases the coupling between the cavity and center, as well as the effective detection efficiency. In fact, with $|\alpha|^2\sim 3$, cooperativities in the strong coupling regime and current detection efficiencies we should be able to achieve a single-shot measurement.

\section{Discussion} 
\label{sec:discussion}

The scheme investigated here uses a dipole-induced transparency to entangle the path of a photon with the spin state of a single nitrogen--vacancy center at cryogenic temperatures. This provides, through subsequent detection of such a photon, a projective measurement on the spin state of the center. The fidelity of this projective measurement forms the key figure of merit in our results. Typically, analyses of these approaches have used significant approximations (e.g. limiting the state space and assuming memoryless scattering distributions) to argue the initial case for their competitiveness, but it is important, before these devices are realized for scalable systems, to determine just how scalable they are. This requires a more in-depth and complete model. In our work here:
\begin{itemize}
	\item The model of the energy level structure of the nitrogen--vacancy center incorporates all ground and optically excited states.
	\item Errors arising from evolution among optically excited states are considered, as well as the ability of the decay path through a meta-stable state to feed back into the correct subspace, when accounting for multiple single-photon pulses.
	\item We characterize the impact of external photon loss and variation in the decay rate to the meta-stable subspace.
	\item With the exception of Section \ref{sec:finite_bandwidth_photons}, scattering probabilities are dependent on the outcomes of preceding measurement pulses.
\end{itemize}
Incorporating these effects allows us to begin to make statements about the applicability of this approach to real, large-scale systems.

Accounting for these additional factors, we expect that the two primary hurdles to the implementation of this scheme are the construction of high-fidelity, narrow-bandwidth ($\sigma_{t}\kappa>>1$) single-photon sources and a cooperativity in the strong-coupling regime (recent work \cite{Janitz15} reported a cooperativity $C\approx 0.2$). The preliminary use of few-photon, weak coherent pulses may circumvent the issue of the single-photon sources, and obviously also increases the effective cooperativity. However, larger scale applications and projective entanglement generation between multiple color centres will require single-photon sources. For the larger applications, charge-state switching and its correction will cause temporal inhomogeneous broadening \cite{Robledo10}, while for entanglement generation, the possibility of stray photons, avoidable only with large cooperativities and high detection efficiencies, degrades the resultant entanglement fidelity. Contemporary difficulties in mind, however, with continuing development we do expect the parameters assumed here for cooperativity ($10$), and source ($60$\%) and detection ($92$\%) efficiencies to be experimentally achievable in the near future; we intend our characterization of the impact of individual error sources to assist experimental efforts to engineer high-fidelity projective operations with this system.

To summarize our view for the immediate future, setting aside the challenges of large-scale applications and projective entanglement generation between nitrogen--vacancy center devices, we envisage that the preliminary use of weak coherent states should allow high-fidelity spin measurements in smaller-scale, contemporary settings. We have estimated an upper bound on the rate of charge-state switching for moderate photon-numbers on the order of $10^{-3}$. This bound, along with the fidelities depicted in Figure \ref{fig:ideal_performance}, suggests that the measurement scheme considered here sees an error rate improvement of, in principle, an order of magnitude over the traditional method of luminescence-detection. While our estimates could be improved with a finer characterization of the ionization rate and the $m_{s}=0$ decay rate through the meta-stable subspace, our results therefore suggest that dipole-induced transparency should provide high fidelity measurement of the spin state of the nitrogen--vacancy center.

\begin{acknowledgments}
We are very grateful to Yuichiro Matsuzaki for helpful comments. This project/publication was made possible through the support of a grant from the John Templeton Foundation. The opinions expressed in this publication are those of the authors and do not necessarily reflect the views of the John Templeton Foundation. KN and MH also acknowledge support from the MEXT KAKENHI Grant-in-Aid for Scientific Research on Innovative Areas “Science of Hybrid Quantum Systems” Grant No. 15H05870, while JS and MT additionally acknowledge the support of the WWTF project ``PhoCluDi'', the FWF project ``SiC-EiC'' (I 3167-N27) and TU Vienna research funds.
\end{acknowledgments}

\bibliography{bibliography.bib}

\appendix
\section{Spin Operators} 
\label{sec:spin_operators}

The Hamiltonians in Equations \ref{eqn:HamGSM}, \ref{eqn:HamESM} and \ref{eqn:HamCoupling} incorporate spin-1 operators $S_{x,y,z}$ and $L_{x,y,z}$. Explicitly, for the $m_{s}$ subsystem these are
\begin{align}
	S_{z} = \left[ \begin{smallmatrix} 1&0&0\\ 0&0&0\\ 0&0&-1 \end{smallmatrix} \right] , \quad S_{x} =& \frac{1}{\sqrt{2}} \left[ \begin{smallmatrix} 0&1&0\\ 1&0&1\\ 0&1&0 \end{smallmatrix} \right], \quad S_{y} = \frac{1}{\sqrt{2}} \left[ \begin{smallmatrix} 0&-i&0\\ i&0&-i\\ 0&i&0 \end{smallmatrix} \right]
\end{align}
so that the raising and lowering operators are given by
\begin{align}
	S_{+} = \sqrt{2} \left[ \begin{smallmatrix} 0&1&0\\ 0&0&1\\ 0&0&0 \end{smallmatrix} \right] , & \qquad S_{-} = \sqrt{2} \left[ \begin{smallmatrix} 0&0&0\\ 1&0&0\\ 0&1&0 \end{smallmatrix} \right] .
\end{align}
Matrices for the $m_{l}$ subsystem are identical.

\section{Equations of Motion}

Now let us derive the equations of motion for our Hamiltonian. This Hamiltonian, in Equations \ref{eqn:HamGSM}, \ref{eqn:HamESM} and \ref{eqn:HamCoupling}, is first expanded to distinguish between photon polarizations ($\sigma_{+}$ and $\sigma_{-}$), so as to preserve orbital angular momentum and enforce polarization selection rules:
\begin{align}
	c L_{x} &\rightarrow c_{\sigma_{+}} L_{+} + c_{\sigma_{-}} L_{-}\\
	c \left( a^{\dagger} + b^{\dagger} \right) &\rightarrow c_{\sigma_{+}} \left( a^{\dagger}_{\sigma_{+}} + b^{\dagger}_{\sigma_{+}} \right) + c_{\sigma_{-}} \left( a^{\dagger}_{\sigma_{-}} + b^{\dagger}_{\sigma_{-}} \right)
\end{align}

Next, exploiting conservation of energy to bind us within the single-excitation subspace, we define composite operators fully characterising the spin and angular momentum states of the system, such that the equations of motion are linear in these operators. Two examples are lowering operators we will arbitrarily call $\hat{C}$ and $\hat{E}$ (for \emph{Cavity}-mode and \emph{Excitation}), corresponding respectively to $c^{\dagger}_{\sigma_{+}}c_{\sigma_{+}}=1,m_{s}=+1,m_{l}=0$ and to $m_{s}=+1,m_{l}=+1$.

\begin{align}
	\hat{C} &= c_{\sigma_{+}} S_{-} \left( S^{2}_{z} + S_{z} \right)/2\\
	\hat{E} &= L_{-} S_{-} \left( L^{2}_{z} + L_{z} \right) \left( S^{2}_{z} + S_{z} \right)/4.
\end{align}

The mean values for $\hat{C}$, $\hat{E}$, and their equivalents for the orthogonal polarization and when $m_{s}\neq +1$ are important because they are directly related to output scattering rates. Finally, we derive Langevin equations for these composite operators of the form
\begin{align}
	 \frac{\partial\hat{C}(t)}{\partial t} &= -i \left[ \left( \omega_{c} + D^{||}_{gs} + \frac{\mu_{B}}{\hbar} g^{||}_{gs} B_{z} - i \frac{\kappa}{2} \right) \hat{C} (t) \right. \nonumber\\
		&\qquad \qquad \left. + \alpha_{\hat{C}} \sqrt{\frac{\kappa}{2\pi}} e^{-ikt} + g \hat{E} (t) \right].
\end{align}

Here $\alpha_{\hat{C}}$ is the amplitude of driving into the $\sigma_{+}$--polarized cavity mode, $k$ is the frequency of this driving, $\omega_{c}$ is the cavity mode frequency, $t$ is the time-dependence of the operators, and other parameters are obtained from the aforementioned Hamiltonians. Performing a Fourier transform gives us a set of linear equations, which can then be solved using the standard methods of numerical linear algebra to obtain the scattering matrix.

\end{document}